\documentclass[conference]{IEEEtran}
\usepackage[utf8]{inputenc}
\usepackage{cite}
\usepackage{amsmath}
\usepackage{amssymb}
\usepackage{amsfonts}
\usepackage{algorithmic}
\usepackage{graphicx}
\usepackage{booktabs}
\usepackage{xcolor}

\usepackage{hyperref}
\hypersetup{pdfborder={0 0 0},colorlinks=true,urlcolor=blue,linkcolor=blue,citecolor=blue,bookmarks=true}
\hypersetup{pdftitle={Implementing Responsible AI: Tensions and Trade-Offs Between Ethics Aspects}}
\hypersetup{pdfauthor={Conrad Sanderson, David Douglas, Qinghua Lu}}

\usepackage[shortlabels]{enumitem}
\usepackage{balance}

\usepackage{inconsolata} 
\usepackage{microtype}
\DisableLigatures[f]{encoding = *, family = * }
\usepackage[absolute]{textpos}

\begin{document}

\title
  {
  {Implementing Responsible~AI:\\Tensions and Trade-Offs Between Ethics Aspects}
  }

\author
  {
  Conrad Sanderson\textsuperscript{{\tiny~}$\dagger\diamond$},
  David Douglas\textsuperscript{{\tiny~}$\ddagger$},
  Qinghua Lu\textsuperscript{{\tiny~}$\dagger$}\\
  ~\\
  \textsuperscript{$\dagger$}{\tiny~}\textit{Data61 / CSIRO, Australia;}~
  \textsuperscript{$\ddagger$}{\tiny~}\textit{Environment / CSIRO, Australia;}~
  \textsuperscript{$\diamond$}{\tiny~}\textit{Griffith University, Australia}
  }

\maketitle

\begin{abstract}

Many sets of ethics principles for responsible AI have been proposed
to allay concerns about misuse and abuse of AI/ML systems.
The underlying aspects of such sets of principles include
privacy, accuracy, fairness, robustness, explainability, and transparency.
However, there are potential tensions between these aspects that pose difficulties
for AI/ML developers seeking to follow these principles.
For example, increasing the accuracy of an AI/ML system may reduce its explainability.
As part of the ongoing effort to operationalise the principles into practice,
in this work we compile and discuss a catalogue of 10 notable tensions, trade-offs and other interactions
between the underlying aspects.
We primarily focus on two-sided interactions, drawing on support spread across a diverse literature.
This catalogue can be helpful in raising awareness
of the possible interactions between aspects of ethics principles,
as well as facilitating well-supported judgements by the designers and developers of AI/ML systems.

\end{abstract}

\vspace{1ex}


\begin{textblock}{13.4}(1.3,14.9)
\hrule
\vspace{1ex}
\noindent
\scriptsize
\textbf{{$^\ast$}~Published in:} International Joint Conference on Neural Networks (IJCNN), 2023. DOI:~\href{https://doi.org/10.1109/IJCNN54540.2023.10191274}{\tt 10.1109/IJCNN54540.2023.10191274}
\end{textblock}

\section{Introduction}

In order to address concerns about potentially negative impacts
of the rapidly growing use of Artificial Intelligence (AI)
and Machine Learning (ML) technologies throughout society~\cite{Whittle_2019},
many governments and organisations have proposed numerous sets of AI ethics principles~\cite{Jobin_2019,Christoforaki_2022}.
An example is given in Fig.~\ref{fig:principles}.

While such principles are well-intended,
a~major issue is that these principles are high-level,
and do not readily provide guidance on how to concretely implement them
within AI/ML systems~\cite{Mikalef_2022,Morley_2020,Morley_2023,Smit_2020}.
Furthermore, attempts at operationalising these principles reveal that many can be in tension
with each other~\cite{Whittlestone_2019,Sanderson_2023}.
This in turn can lead to suboptimal outcomes, 
where the designers and developers of AI/ML systems
may haphazardly resolve tensions by simply selecting one principle (or its underlying aspect) to be dominant,
rather than devising and incorporating well-balanced trade-offs between the principles in tension 
where the associated risks and benefits are gauged more thoroughly~\cite{Whittlestone_2019}.

\begin{figure}[!b]
\vspace{-2ex}
\centering
\fbox{%
\centering
\begin{minipage}{0.96\columnwidth}
\normalsize
\begin{enumerate}[{\tt 1.},leftmargin=*]
\small
\itemsep=0.5ex
\vspace{1ex}

\item
{\bf Human, Social \& Environmental Wellbeing.}
AI systems should benefit individuals, society, and the environment.

\item
{\bf Human-centred Values.}
AI systems should respect human rights, diversity, and the autonomy of individuals.

\item
{\bf Fairness.}
AI systems should be inclusive and accessible,
and should not involve or result in unfair discrimination against individuals or groups.

\item
{\bf Privacy Protection \& Security.}
AI systems should respect and uphold privacy rights and regulations, and ensure the security of data.

\item
{\bf Reliability \& Safety.}
AI systems should reliably operate in the context of their intended purpose throughout their lifecycle.

\item
{\bf Transparency \& Explainability.}
There should be transparency and responsible disclosure
allowing people to know when an AI system is engaging with them
and/or when they are significantly impacted by an AI system.

\item
{\bf Contestability.}
When an AI system significantly impacts a person, group or environment,
there should be a timely process to allow people to challenge the use or output of the system.

\item
{\bf Accountability.}
Human oversight of AI systems should be enabled;
those responsible for the various phases of the AI system lifecycle
should be identifiable and accountable for the outcomes of the system.

\vspace{1ex}
\end{enumerate}
\end{minipage}
} 
\caption
  {
  Summarised form of the high-level AI ethics principles 
  proposed by the Australian Government~\cite{AusAIprinciples_2019}.
  }
\label{fig:principles}
\end{figure}

Although the available sets of AI ethics principles have notable differences in wording,
a set of underlying common themes and aspects can be identified within them~\cite{Fjeld_2020,Jobin_2019}.
These common aspects typically include privacy, accuracy, fairness, robustness, explainability, 
transparency, and accountability.
Tensions can exist between such aspects which require trade-offs to be made between them;
for example, increasing privacy may come at the cost of decreasing accuracy (a negative interaction).
However, interactions between the aspects are not always in the forms of trade-offs:
increasing robustness, for example, may increase accuracy in certain circumstances (a~context-dependent interaction).
Furthermore, interactions may also benefit all the involved aspects (positive interactions).

There are many observed interactions (positive, negative, and context-dependent)
between various two- and three-sided combinations of the common aspects of AI ethics principles.
However, descriptions of such interactions are spread across a diverse and disparate literature.
This in turn may contribute to a lack of awareness among the designers and developers of AI/ML systems
about the wide range and nature of the possible interactions between the underlying aspects.

In this work we aim to address the above problem in a two-part fashion.
We first gather and provide definitions of the underlying aspects present in AI ethics principles, 
shedding light on multiple (possibly semi-compatible) definitions of each aspect.
Building on~\cite{Lee_2020}, the various tensions, conflicts, trade-offs
and other interactions between the aspects that can occur in practice are then elucidated and discussed,
as supported by the available literature.
Such an explicit catalogue can be helpful
in raising awareness of the possible interactions,
as well as facilitating well-supported judgements for implementing \textit{responsible AI} 
by the designers and developers of AI/ML systems.

The structure of this paper is as follows.
The descriptions of each aspect are provided in Section~\ref{sec:aspects}.
The interactions between the aspects are explored in Section~\ref{sec:interactions}.
Concluding remarks and future avenues of research are given in Section~\ref{sec:conclusion}.

\section{Underlying Aspects}
\label{sec:aspects}

Across the many proposed sets of AI ethics principles, 
we focus on six underlying aspects:
(a)~accuracy,
(b)~robustness,
(c)~fairness,
(d)~privacy,
(e)~explainability,
(f)~transparency.
An~overview of each of these aspects is given below.

The selection of these aspects is driven by the availability of literature
that covers interactions between at least two such aspects.
We have elected not to explicitly cover the \textit{accountability} aspect,
due to it not being easily quantifiable,
as well as its close association with the \textit{transparency} aspect~\cite{diakopoulos2020}.

We note that while the names of the aspects may not directly match the wording used
within a given set of AI ethics principles,
the names can be easily mapped to the corresponding wording.
For example, \textit{robustness} can be mapped to \textit{reliability}
within the set of high-level AI ethics principles shown in Fig.~\ref{fig:principles}
with a high degree of fidelity.
Furthermore, based on the definitions available in the literature,
we have elected to place related and/or semi-compatible definitions under one umbrella.

\subsection{Accuracy}

Given data comprising instances of samples along with associated ground-truth labels,
\textit{accuracy}, in its simplest form, can be defined as the ratio of correct classifications
to the total number of classifications made by an AI/ML model.
The act of classification in this case is defined
as the process of mapping a given sample to one label out of a set of pre-defined labels.
Accuracy is typically measured on a hold-out dataset:
data that is known and available, but is not used for training the AI/ML models.
This straightforward measurement approach aims
to evaluate the generalisation ability of the models under known (normal) conditions~\cite{Bishop_2006}.

In the case of binary classifications,
where the possible decisions (outputs from an AI/ML model) are \textit{true} and \textit{false},
more complex forms of accuracy are used,
such as \textit{recall}, \textit{precision}, \textit{F1~score},
and \textit{area under curve}~\cite{Bradley_1997,Bishop_2006}.
There are also task-specific measures of accuracy
where the notions of \textit{correct} and \textit{incorrect} are less clearly defined,
such as in language translation~\cite{Papineni_2002}.

Specialisations of accuracy can also be denoted as \textit{overall accuracy} and \textit{group accuracy}.
In the former, the entire hold-out dataset is used,
where each sample is treated individually and has equal contribution to the final measurement;
this is the traditional definition of accuracy~\cite{Bishop_2006}.
In the latter, accuracy is calculated for a pre-defined group in the hold-out dataset,
where the group membership is defined by one or more characteristics, such as gender and ethnicity.
Within this work we use the term \textit{accuracy} to mean \textit{overall accuracy},
unless specified otherwise.

\subsection{Robustness}
\vspace{-0.25ex}

The robustness of an AI/ML system is its ability to perform as expected
under various conditions~\cite{Cooper_2022}.
A classical definition of robustness within machine learning
is the ability to maintain a high degree of accuracy
under conditions that may have not been observed in the training dataset~\cite{Wong_2012}.
A~related definition of robustness is the capacity to maintain high accuracy
under a wide range of conditions~\cite{Zhang_2018}.

A more recent definition of robustness is resistance to adversarial attacks,
which may occur during training and/or usage of an AI/ML model~\cite{Alshemali_2020,Akhtar_2021,Szegedy_2014,Li_2022,Xu_2020}.
In both cases, data is deliberately altered or constructed 
so that the model is induced to perform unreliably (eg.~reduction in accuracy),
or is biased towards producing results favoured by an attacker (eg.~reduction in accuracy for a specific group).
Robustness can also be defined as resiliency to label noise,
where the class labels within training data
are deliberately or inherently inaccurate~\cite{Frenay_2014,Vahdat_2017,Hendrycks_2019}.

The nature of the attacks can be placed into three broad categories:
so-called white-box, black-box, and grey-box attacks~\cite{Akhtar_2021}.
In the white-box case, the attacks can be dependent on the type and underlying characteristics of an AI/ML model
(cf.~transparency in Section~\ref{sec:transparency}).
In the black-box case, the attacks may not have access to such details,
and hence may also be probing in nature, designed to glean information
about the capabilities and limitations of an AI/ML system.
In the grey-box case, the attacker has partial knowledge of the system.
An instructive example of adversarial attacks in the context of deep neural networks is shown in~\cite{Dezfooli_2016}.

\vspace{-0.25ex}
\subsection{Fairness}
\vspace{-0.25ex}

Within the context of machine learning, fairness can considered as the opposite of bias,
where bias can be construed as favouring a particular characteristic (eg.~gender) or group.
In a simplified view, two conflicting definitions of fairness exist~\cite{Friedler_2021,Yeom_2021}:
\textbf{(i)} individual fairness
(eg.~fairness as equality of treatment across individuals),
and
\textbf{(ii)} group fairness (eg.~fairness as equality of outcomes across demographic groups).

Under definition (i),
an AI/ML model that obtains the lowest error rates
for the largest percentage of individuals is considered the fairest
(ie.~aims for maximising overall accuracy).
Under definition (ii), an AI/ML model that obtains similar error rates across groups is the fairest 
(ie.~aims for minimising differences between group accuracies).
Representative fairness metrics include demographic parity, equality of odds, and equality of opportunity~\cite{Du_2021}.

\vspace{-0.25ex}
\subsection{Privacy}
\vspace{-0.25ex}

Distinct from protection of sensitive data via access control means (eg.~via passwords),
privacy in the context of AI/ML systems can be defined as preventing inadvertent or deliberate
leakage of sensitive data via algorithmic means. 
A~straightforward approach is to remove personally identifiable information from training data (anonymisation).
However, given a sufficient amount of data from external sources,
it may still be possible to de-anonymise individuals~\cite{Narayanan_2008, henriksen-bulmer_2016}. 

Approaches aiming to counteract de-anonymisation include differential privacy~\cite{Jagielski_2019,Yu_2019}, 
which can be roughly described as adding noise to input data,
model parameters, and/or output data.
Other techniques aiming towards protection of sensitive data include 
federated learning~\cite{TianLi_2020,Wilson_2022}
and
homomorphic encryption~\cite{Acar_2018,Jain_2022}.
In federated learning an AI/ML model is trained using multiple decentralised computers;
each computer uses only a subset of the training dataset and has no access to the other subsets.
In homomorphic encryption, computations are performed on encrypted data without prior decryption of the data,
thereby allowing data processing to be done by third parties (eg.~cloud services).
To protect data privacy in model exchanges and aggregations,
federated learning can be implemented via secure multi-party computation
in conjunction with homomorphic encryption~\cite{Keller_2022_v2}.

\subsection{Explainability}
\label{sec:explainability}

Explainability, in the strictest sense,
can be defined as the provision of a mathematical explanation
of how an AI/ML system processes given input data to produce a specific output
(eg.~the processing steps of a neural network).
However, given the large scale and complexity of modern AI/ML pipelines
(eg.~deep neural networks with millions of parameters),
complete explanations are very difficult to interpret by humans~\cite{Adadi_2018,Arrieta_2020}.
Systems with these characteristics are often informally referred to as
``black boxes''~\cite{Adadi_2018,Rudin_2019}.

Various approaches are available to provide partial explainability
that allow forms of interpretability
(ie.~reasoning behind a given output from an AI/ML system)~\cite{Adadi_2018,Arrieta_2020}.
This includes text-based and visual explanations~\cite{Ribeiro_2016},
as well as post-hoc methods,
such as the creation of simplified surrogate models that are easier to interpret~\cite{Arrieta_2020,Sanderson_2023}.
The required degree of explainability may depend on the intended audience (eg.~developers, end users)
as well as their background, culture, and preferences~\cite{Arrieta_2020,Sanderson_2023}.

\subsection{Transparency}
\label{sec:transparency}

Transparency can be defined as the provision of partial or full details of the AI/ML system,
including the type of machine learning algorithm employed,
the datasets used for training and evaluation 
as well as the training procedure (including data pre-processing),
and associated performance metrics (eg.~accuracy rates on the evaluation dataset).
The degree of transparency may be restricted by commercial concerns 
(eg.~availability of source code, access to datasets)~\cite{Burrell_2016}.
Transparency can also be defined as allowing users to be aware
that they are (directly or indirectly) interacting with an AI/ML system~\cite{AusAIprinciples_2019}.

In both definitions, transparency can be considered as a necessary component of accountability,
which aims to identify the persons and/or organisations responsible for specific components
of the overall AI/ML system~\cite{diakopoulos2020}.

The notion of explainability is sometimes conflated with the term transparency 
by designers and developers of AI/ML systems~\cite{Sanderson_2023},
where the latter term is used as a synonym for the former term (cf.~Section~\ref{sec:explainability}).

\section{Interactions between Aspects}
\label{sec:interactions}

We describe and discuss the following interactions,
based on support available in the literature:

\vspace{1ex}
\setlength{\tabcolsep}{0.5ex}
\begin{tabular}{cl}
(a) & Accuracy vs.~Robustness \\
(b) & Accuracy vs.~Fairness \\
(c) & Accuracy vs.~Privacy \\
(d) & Accuracy vs.~Explainability \\
(e) & Fairness vs.~Robustness \\
(f\hspace{0.2ex}) & Fairness vs.~Privacy \\
(g) & Fairness vs.~Transparency \\
(h) & Privacy vs.~Robustness \\
(i) & Transparency vs.~Explainability \\
(\hspace{0.2ex}j) & Transparency vs.~Privacy and Robustness
\end{tabular}

\subsection{Accuracy vs.~Robustness}

Machine learning models that achieve high accuracy
within the conditions defined by the training dataset (representing expected usage)
can quickly degrade in performance when they do not explicitly take
into account that the conditions during deployment can be significantly different
due to (potentially foreseeable) changes in data acquisition circumstances~\cite{Wong_2012,Zhang_2018}.

The accuracy of AI/ML systems can also degrade over time due to natural data drift~\cite{Lewis_2022}.
Periodic recalibration and/or retraining of AI/ML systems may be necessary
to ensure the system is working accurately in changed conditions~\cite{Lewis_2022,Sanderson_2023}.
Such an approach is also known as \textit{algorithmic maintenance}~\cite{Lee_2020}.
However, blindly updating training data without thoroughly testing the retrained system 
may result in instabilities and/or unintended changes in operation~\cite{Sculley_2015}.

Deliberate adversarial attacks during usage of an AI/ML model may use data that is designed to fall outside
of the normal conditions that the model was originally trained to handle~\cite{Madry_2018,Akhtar_2021,Li_2022,Xu_2020}.
Increasing robustness against adversarial attacks may involve
augmenting the training dataset (eg.~with perturbed versions of the original data),
resulting in a larger training dataset.
Related approaches include the generation of additional training samples via convex linear combinations of existing samples,
and taking into account during model training the immediate neighbourhood of each training sample
in addition to the sample itself~\cite{Akhtar_2021,Li_2022,Xu_2020}.

When robustness is defined as resiliency to label noise,
a positive interaction between accuracy and robustness is likely to be present~\cite{Vahdat_2017,Hendrycks_2019}.
When robustness is defined as resistance to adversarial attacks,
the link between accuracy and robustness appears to have two facets.
For small training datasets, taking into account robustness may increase accuracy;
in contrast, for larger training datasets, taking into account robustness is likely to decrease accuracy
\cite{Tsipras_2019,Su_2018,Lei_2019,Zhang_2019}.
For small amounts of training data, the resultant models are likely to be overfitting~\cite{Bishop_2006},
and hence incorporating robustness can act as a form of regularisation.
With larger amounts of training data, the resultant models are already well-fitting
and do not require further regularisation.
The reduction in accuracy due to adversarial attacks
can be somewhat controlled using specifically crafted ML models~\cite{Pinot_2022}.

\subsection{Accuracy vs.~Fairness}
\label{sec:accuracy_vs_fairness}
\vspace{-0.26ex}

Recent literature indicates that incorporating fairness provisions
(under all definitions of fairness) never improves accuracy, and typically hinders it
\cite{Benz_2021,Wadsworth_2018,Jagielski_2019,Friedler_2019}.
Overall, the interaction can be summarised as negative.
For example, with fairness defined as group fairness,
as fairness increases, overall accuracy is expected to decrease~\cite{Jagielski_2019}.

It has also been observed that accuracy can be group-dependent
even if the training dataset is balanced (ie.~similar number of training samples for each group).
Depending on the nature of a given group,
accuracy for that group may be inherently lower than for other groups~\cite{Benz_2021}.
A notable example of this disparity is the relatively low accuracy of face recognition
for females with darker skin complexion~\cite{Buolamwini_2018}.

In scenarios involving medical applications,
developers of an AI/ML system may need to restrict its applicability to a specific ethnic group,
due to limitations in the availability of sufficient high-quality training data.
Without such restriction, the system is likely to provide inaccurate outputs~\cite{Sanderson_2023}.

\vspace{-0.26ex}
\subsection{Accuracy vs.~Privacy}
\vspace{-0.26ex}

Accuracy is typically decreased when differential privacy is employed,
with the decrease being typically minor
\cite{Shokri_2015,Phan_2020,Jayaraman_2019}.
However, the decrease in accuracy can be considerably greater
for under-represented groups (eg.~minorities)~\cite{Bagdasaryan_2019},
which in turn can affect group fairness.

Accuracy can be decreased when privacy considerations drive the minimisation of data acquisition,
in order to reduce the amount of personally identifiable information that could be potentially exposed~\cite{Sanderson_2023}.
The exposure can be due to security lapses or adversarial attacks.

In cases where the use of federated learning is appropriate for addressing privacy concerns,
accuracy may decrease if there are significant differences in the nature of the separate datasets
held in the separate computing nodes~\cite{QiangYang_2019,TianLi_2020}.
However, when AI/ML model training time is limited,
the parallelised nature of federated learning can lead to increased accuracy.
In~contrast to using a single computing node,
federated learning is quicker as multiple computing nodes are employed,
thereby processing more training data within an allotted time budget~\cite{Sanderson_2023}.
Larger amounts of training data typically lead to more accurate AI/ML models~\cite{Bishop_2006}.

\vspace{-0.26ex}
\subsection{Accuracy vs.~Explainability}
\vspace{-0.26ex}

In cases where explainability is required
(for internal development purposes and/or to provide reasoning for end users),
designers and developers of AI/ML systems may choose to 
employ methods/models that are easier to interpret but are less accurate,
rather than more accurate approaches where it is more
difficult to convey the reasoning behind their outputs~\cite{Rudin_2019,Sanderson_2023}.
However, accuracy and explainability are not necessarily mutually exclusive,
as it is possible to devise a trade-off between them
by varying model complexity~\cite{Petkovic_2023}.

Simplified (and hence more readily interpretable) models can be also used 
purely for the purpose of driving explanations (ie.~surrogate models),
rather than as replacements of the original (non-interpretable) models~\cite{Bastani_2017}.
In this case the explanations obtained via surrogate models do not affect the accuracy of the original model.
However, there is a question on the fidelity of the surrogate model,
in the sense that the obtained explanations may not reflect the underlying processing
within the original model~\cite{Adadi_2018}.

\vspace{-0.26ex}
\subsection{Fairness vs.~Robustness}
\label{sec:fairness_vs_robustness}
\vspace{-0.26ex}

To achieve robust functionality, the range of allowable conditions may need to be constrained,
which in turn may be in tension with fairness.
Using an AI/ML system beyond what the training dataset covers (eg.~people with specific ethnicity)
may result in unreliable operation~\cite{Sanderson_2023}.

When explicitly taking into account robustness against adversarial attacks,
group fairness can be reduced, even when the training dataset is balanced across the classes.
Furthermore, inherent differences in accuracy across groups can be exacerbated~\cite{Benz_2021}.

Within AI/ML approaches focused on text classification, 
robustness to targeted word substitutions can help
with improving both individual fairness~\cite{Yurochkin_2020} and group fairness~\cite{Garg_2019}.
The targeted substitutions typically relate to words associated with a protected subgroup.
Examples include robustness to changes in descriptions of sentence toxicity~\cite{Garg_2019}
and changes in gender~\cite{Yurochkin_2020}.

\vspace{-0.26ex}
\subsection{Fairness vs.~Privacy}
\vspace{-0.26ex}

Fairness tends to be reduced as a byproduct of employing differential privacy,
as exemplified in~\cite{Bagdasaryan_2019,Jagielski_2019,Cummings_2019}.
Furthermore, non-fairness can be exacerbated if the original AI/ML model produces non-fair outcomes,
with the decrease in fairness more pronounced for minority groups with inadequate representation~\cite{Bagdasaryan_2019}.
However, it is possible to ameliorate this effect
and achieve approximate fairness under the constraints of differential privacy
with specifically constructed machine learning models~\cite{Cummings_2019}.

\vspace{-0.26ex}
\subsection{Fairness vs.~Transparency}
\vspace{-0.26ex}

Through increasing transparency of an AI/ML system it may become apparent
to the designers or the end users that the system is unfair~\cite{Arrieta_2020},
thereby possibly leading to internal or external pressure to increase fairness.
Furthermore, if the designers of an AI/ML system are required to demonstrate that the system is fair,
an effective means for doing so is to increase transparency (and possibly explainability).

The appropriate definition of fairness for an AI/ML system will depend on the context of its use~\cite{Veale_2017}.
Disclosing how designers have understood and implemented fairness
will assist the end users of an AI/ML system in evaluating whether it is likely to produce fair outcomes.
This is particularly important where end users are accountable for the
decisions made using an AI/ML system~\cite{Veale_2018}. 

A possible tension between fairness and transparency
is the potential for malicious or mischievous actors
to exploit an AI/ML system to their own advantage
based on the information disclosed about the system~\cite{Kroll_2017, Veale_2018}.
In such cases, transparency may need to be limited to protect the fairness of the system.
The designers should consider to whom the AI/ML system should be transparent,
and what information should be provided~\cite{diakopoulos2020}.
A similar tension is created if aspects of an AI/ML system are protected as trade secrets~\cite{Burrell_2016}.
To maintain accountability where public transparency is impossible,
the AI/ML system may be made transparent to a trusted auditor for review~\cite{Pasquale_2015}
(cf. Section~\ref{sec:transparency_vs_privacy_and_robustness}).

\vspace{-0.26ex}
\subsection{Privacy vs.~Robustness}
\vspace{-0.26ex}

At a high level, privacy can be linked with robustness through
the overarching aim of preventing the misuse of AI/ML systems.
This is reflected in the literature,
where both differential privacy and robustness to adversarial attacks
can be incorporated into the same machine learning approach~\cite{Phan_2020}.
Furthermore, the link between privacy and robustness can be exploited
to create a foundation for building AI/ML methods
that explicitly takes into account both requirements at the same time~\cite{Lecuyer_2019}.

\vspace{-0.26ex}
\subsection{Transparency vs.~Explainability}
\vspace{-0.26ex}

Explainability can help with transparency,
in the sense that having a documented understanding of the AI/ML pipeline (necessary for explainability)
can be co-opted to provide abridged details (suitable for disclosure) about the AI/ML
system~\cite{diakopoulos2020}. Explainability should be understood as being targeted towards
a particular audience~\cite{Arrieta_2020}.
These audiences will depend on the context within which an AI/ML system is used. 
Example audiences may be the end users of an AI/ML system,
those affected by  decisions made by an AI/ML system, or regulators.
The details of an AI/ML system  should be presented in formats
that are easily understood and used by their target audiences~\cite{diakopoulos2020}.

\vspace{-0.26ex}
\subsection{Transparency vs.~Privacy and Robustness}
\label{sec:transparency_vs_privacy_and_robustness}
\vspace{-0.26ex}

Increasing transparency can negatively affect both robustness and privacy,
in the sense that revealing details about an AI/ML system,
such as its neural network architecture and training procedure (including associated datasets),
can facilitate the potential re-identification (de-anonymisation) of individuals~\cite{Narayanan_2008,Rocher_2019},
as well as to perform targeted adversarial attacks in order to cause malfunction.

On the surface it may appear that opaque systems
are more difficult to analyse and hence more difficult to exploit~\cite{Papernot_2017}.
However, this view has parallels with the traditional security approach in proprietary (closed-source) software systems,
which has been pejoratively referred to as \textit{security through obscurity}~\cite{Anderson_2001,Mercuri_2003}
and its efficacy questioned in contrast to open-source software~\cite{Mercuri_2005,Hoepman_2007}.

Given these opposing viewpoints,
it may be preferable to seek a balance between the two extremes of transparency
(ie.~full transparency vs.~no transparency),
such as providing partial transparency to trusted parties,
with the degree of transparency dependent on the degree of trust~\cite{Pasquale_2015,Young_2019}.

Transparency can be also hampered through the use of proprietary datasets for training AI/ML systems,
where the owner of the dataset may prevent disclosure that the dataset is used
or prevent sharing it with third parties,
due to commercial sensitivities and/or privacy considerations~\cite{Innerarity_2021,Sanderson_2023}.

\section{Concluding Remarks}
\label{sec:conclusion}

Various sets of AI ethics principles typically have a common set of underlying aspects,
which includes accuracy, fairness, robustness, privacy, explainability, and transparency.
These aspects may interact with each other, and require designers and developers to make trade-offs between them.
While some tensions between these aspects may be generally known (such as the trade-off between accuracy and privacy),
the wider gamut of observed interactions between the aspects is spread across a diverse literature
and not well-known as a whole.

In order to facilitate the operationalisation of AI ethics principles,
in this work we have compiled and elucidated a catalogue of 10 notable interactions between the underlying aspects,
primarily focusing on two-sided interactions for which there is support in the literature.
While we have also covered a three-sided interaction (transparency vs. privacy and robustness),
future avenues of research include the exploration of other possible three-sided interactions,
such as
accuracy vs. robustness vs. fairness~\cite{Benz_2021b,Benz_2021},
and
accuracy vs. privacy vs. fairness~\cite{Gu_2022,Kesari_2022}.

Recognising the tensions and other interactions between common aspects of AI ethics
principles is an important step towards operationalising these principles. 
The catalogue presented here can be helpful in raising awareness of the possible interactions between these aspects,
as well as facilitating well-supported judgements by the designers and developers of AI/ML systems in addressing them.
In follow-up work we examine several notable approaches for resolving the tensions~\cite{Sanderson_2024}.

%
%
%


\bibliographystyle{ieee_mod}
\bibliography{references}

\end{document}